\documentclass{pasj00}
\draft

\begin{document}
\SetRunningHead{Author(s) in page-head}{Running Head}

\title{Indications of M-dwarf Deficits in the Halo and Thick Disk of the Galaxy}

 \author{
   Mihoko~\textsc{Konishi}\altaffilmark{1},
   Hiroshi~\textsc{Shibai}\altaffilmark{1},
   Takahiro~\textsc{Sumi}\altaffilmark{1},
   Misato~\textsc{Fukagawa}\altaffilmark{1},
   Taro~\textsc{Matsuo}\altaffilmark{2},
   Matthias~S.~\textsc{Samland}\altaffilmark{3},
   Kodai~\textsc{Yamamoto}\altaffilmark{2},
   Jun~\textsc{Sudo}\altaffilmark{1},
   Yoichi~\textsc{Itoh}\altaffilmark{4},
   Nobuo~\textsc{Arimoto}\altaffilmark{5,6},
   Masaru~\textsc{Kajisawa}\altaffilmark{7},
   Lyu~\textsc{Abe}\altaffilmark{8},
   Wolfgang~\textsc{Brandner}\altaffilmark{9},
   Timothy~D.~\textsc{Brandt}\altaffilmark{10},
   Joseph~\textsc{Carson}\altaffilmark{11},
   Thayne~\textsc{Currie}\altaffilmark{12},
   Sebastian~E.~\textsc{Egner}\altaffilmark{5},
   Markus~\textsc{Feldt}\altaffilmark{9},
   Miwa~\textsc{Goto}\altaffilmark{13},
   Carol~A.~\textsc{Grady}\altaffilmark{14, 15},
   Olivier~\textsc{Guyon}\altaffilmark{5},
   Jun~\textsc{Hashimoto}\altaffilmark{16},
   Yutaka~\textsc{Hayano}\altaffilmark{5},
   Masahiko~\textsc{Hayashi}\altaffilmark{17},
   Saeko~S.~\textsc{Hayashi}\altaffilmark{5},
   Thomas~\textsc{Henning}\altaffilmark{9},
   Klaus~W.~\textsc{Hodapp}\altaffilmark{18},
   Miki~\textsc{Ishii}\altaffilmark{5},
   Masanori~\textsc{Iye}\altaffilmark{17},
   Markus~\textsc{Janson}\altaffilmark{19},
   Ryo~\textsc{Kandori}\altaffilmark{17},
   Gillian~R.~\textsc{Knapp}\altaffilmark{10},
   Tomoyuki~\textsc{Kudo}\altaffilmark{5},
   Nobuhiko~\textsc{Kusakabe}\altaffilmark{17},
   Masayuki~\textsc{Kuzuhara}\altaffilmark{20, 17, 21},
   Jungmi~\textsc{Kwon}\altaffilmark{17, 21},
   Michael~W.~\textsc{McElwain}\altaffilmark{15},
   Shoken~\textsc{Miyama}\altaffilmark{22},
   Jun-Ichi~\textsc{Morino}\altaffilmark{17},
   Amaya~\textsc{Moro-Mart\'{i}n}\altaffilmark{23},
   Tetsuo~\textsc{Nishimura}\altaffilmark{5},
   Tae-Soo~\textsc{Pyo}\altaffilmark{5},
   Eugene~\textsc{Serabyn}\altaffilmark{24},
   Takuya~\textsc{Suenaga}\altaffilmark{6},
   Hiroshi~\textsc{Suto}\altaffilmark{17},
   Ryuji~\textsc{Suzuki}\altaffilmark{17},
   Yasuhiro~H.~\textsc{Takahashi}\altaffilmark{21, 17},
   Hideki~\textsc{Takami}\altaffilmark{17},
   Naruhisa~\textsc{Takato}\altaffilmark{5},
   Hiroshi~\textsc{Terada}\altaffilmark{5},
   Christian~\textsc{Thalmann}\altaffilmark{25}, 
   Daigo~\textsc{Tomono}\altaffilmark{5},
   Edwin~L.~\textsc{Turner}\altaffilmark{10},
   Tomonori~\textsc{Usuda}\altaffilmark{17},
   Makoto~\textsc{Watanabe}\altaffilmark{26},
   John~P.~\textsc{Wisniewski}\altaffilmark{16},
   Toru~\textsc{Yamada}\altaffilmark{27},
   and
   Motohide~\textsc{Tamura}\altaffilmark{21,17}}
 \altaffiltext{1}{Department of Earth and Space Science, Graduate School of Science, Osaka University, 1-1 Machikaneyama, Toyonaka, Osaka 560-0043, Japan}\email{konishi@iral.ess.sci.osaka-u.ac.jp}
 \altaffiltext{2}{Department of Astronomy, Faculty of Science, Kyoto University, Kitashirakawa-Oiwake-cho, Sakyo-ku, Kyoto 606-8502, Japan}
 \altaffiltext{3}{Department of Physics and Astronomy, Heidelberg University, Seminarstra\ss e 2 69117 Heidelberg, Germany}
 \altaffiltext{4}{Nishi-Harima Astronomical Observatory, 407-2 Nishigaichi, Sayo-cho, Sayo-gun, Hyogo 679-5313, Japan}
 \altaffiltext{5}{Subaru Telescope, 650 North A'ohoku Place, Hilo, HI 96720, USA}
 \altaffiltext{6}{The Graduate University for Advanced Studies (SOKENDAI), 2-21-1 Osawa, Mitaka, Tokyo 181-8588, Japan}
 \altaffiltext{7}{Research Center for Space and Cosmic Evolution, Ehime University, Bunkyo-cho 2-5, Matsuyama, Ehime, 790-8577, Japan}
 \altaffiltext{8}{Laboratoire Lagrange (UMR 7293), Universit\'{e} de Nice-Sophia Antipolis, CNRS, Observatoire de la C\^{o}te d'Azur, 28 avenue Valrose, 06108 Nice Cedex 2, France}
 \altaffiltext{9}{Max Planck Institute for Astronomy, K\"{o}nigstuhl 17, D-69117 Heidelberg, Germany}
 \altaffiltext{10}{Department of Astrophysical Science, Princeton University, Peyton Hall, Ivy Lane, Princeton, NJ 08544, USA}
 \altaffiltext{11}{Department of Physics and Astronomy, College of Charleston, 58 Coming St., Charleston, SC 29424, USA}
 \altaffiltext{12}{Department of Astronomy and Astrophysics, University of Toronto, 50 St. George Street, Toronto, ON, Canada}
 \altaffiltext{13}{Universit\"{a}ts-Sternwarte M\"{u}nchen, Ludwig-Maximilians-Universit\"{a}t, Scheinerstr. 1, 81679 M\"{u}nchen, Germany}
 \altaffiltext{14}{Eureka Scientific, 2452 Delmer, Suite 100, Oakland CA 96002, USA}
 \altaffiltext{15}{ExoPlanets and Stellar Astrophysics Laboratory, Code 667, Goddard Space Flight Center, Greenbelt, MD 20771, USA}
 \altaffiltext{16}{H.L. Dodge Department of Physics and Astronomy, University of Oklahoma, 440 W Brooks St Norman, OK 73019, USA}
 \altaffiltext{17}{National Astronomical Observatory of Japan, 2-21-1, Osawa, Mitaka, Tokyo, 181-8588, Japan}
 \altaffiltext{18}{Institute for Astronomy, University of Hawaii, 640 N. A'ohoku Place, Hilo, HI 96720, USA}
 \altaffiltext{19}{Astrophysics Research Center, Queen's University Belfast, BT7 1NN, Northern Ireland, UK}
 \altaffiltext{20}{Department of Earth and Planetary Sciences, Tokyo Institute of Technology, 2-12-1 Ookayama, Meguro-ku, Tokyo 152-8551, Japan}
 \altaffiltext{21}{Department of Earth and Planetary Science, The University of Tokyo, 7-3-1, Hongo, Bunkyo-ku, Tokyo 113-0033, Japan}
 \altaffiltext{22}{Hiroshima University, 1-3-1, Kagamiyama, Higashi-Hiroshima, Hiroshima 739-8526, Japan} 
 \altaffiltext{23}{Department of Astrophysics, CAB-CSIC/INTA, 28850 Torrejn de Ardoz, Madrid, Spain}
 \altaffiltext{24}{Jet Propulsion Laboratory, California Institute of Technology, Pasadena, CA, 91109, USA}
 \altaffiltext{25}{Institute for Astronomy, ETH Zurich, Wolfgang-Pauli-Strasse 27, 8093 Zurich, Switzerland}
 \altaffiltext{26}{Department of Cosmosciences, Hokkaido University, Kita-ku, Sapporo, Hokkaido 060-0810, Japan}
 \altaffiltext{27}{Astronomical Institute, Tohoku University, Aoba-ku, Sendai, Miyagi 980-8578, Japan}


\maketitle

\begin{abstract}
We compared the number of faint stars detected in deep survey fields with the current stellar distribution model of the Galaxy and found that the detected number in the $H$ band is significantly smaller than the predicted number. This indicates that M-dwarfs, the major component, are fewer in the halo and the thick disk. We used archived data of several surveys in both the north and south field of GOODS (Great Observatories Origins Deep Survey), MODS in GOODS-N, and ERS and CANDELS in GOODS-S. The number density of M-dwarfs in the halo has to be $20 \pm 13 \%$ relative to that in the solar vicinity, in order for the detected number of stars fainter than 20.5 mag in the $H$ band to match with the predicted value from the model. In the thick disk, the number density of M-dwarfs must be reduced ($52 \pm 13 \%$) or the scale height must be decreased ($\sim600$~pc). Alternatively, overall fractions of the halo and thick disks can be significantly reduced to achieve the same effect, because our sample mainly consists of faint M-dwarfs. Our results imply that the M-dwarf population in regions distant from the Galactic plane is significantly smaller than previously thought. We then discussed the implications this has on the suitability of the model predictions for the prediction of non-companion faint stars in direct imaging extrasolar planet surveys by using the best-fit number densities.
\end{abstract}

\section{Introduction} \label{sec:intro}

Direct imaging surveys of extrasolar planets are conducted by using large telescopes (e.g. \cite{bowler13, chauvin14, nielsen13, yamamoto13}). Many planet candidates, whose magnitudes reach around 22~mag in the $H$ band (1.6~\micron) and $K$ band (2.2~\micron), were detected. However, only about 10 extrasolar planets have been confirmed\footnote{Extrasolar Planets Encyclopaedia, http://exoplanet.eu/ (as at Feb. 2014)}. In other words, most of the detected candidates are stars that are not companions, located in the same line of sight, despite the primary goal of detecting extrasolar planets. It is therefore important to estimate the number of unbound stars in a given field of view (FoV).

The mean number of such stars can be estimated using a stellar distribution model of the Galaxy. The standard stellar distribution model was proposed based on optical observations (e.g. \cite{bahcall80, gilmore83}). \citet{juric08} modified parameters of this model, such as the scale height and length, using the SDSS (Sloan Digital Sky Survey) up to 22~$AB$~mag in the $r$ band (6231~\AA), in order to match with deeper optical star counts. The same distribution profiles of the aforementioned model are applicable to the infrared star counts (e.g. \cite{jones81, wainscoat92}). Recently, some studies were conducted to optimize parameters of this model to match with the space telescope catalogues up to around 15~mag in the near-infrared (NIR) band (e.g. \cite{robin03, chang12, polido13, czekaj14}). \citet{nakajima00} conducted deep imaging observations directed toward the north Galactic pole in the $J$ band using Subaru Telescope, and their findings showed that observations were consistent with the model's predictions, including M-dwarfs and brown dwarfs (L- and T-type stars). However, they could not quantitatively discuss the number densities of these stars due to the small sample size.

The stellar distribution model of the Galaxy was evaluated only up to 15~mag in the NIR band, although detected stars are generally fainter for extrasolar planet surveys. There is some doubt as to whether this model can be used to estimate the number of such faint stars. It is highly probable that these stars are distant late-type dwarfs. Actually, it is difficult to establish the number densities of these dwarfs outside the solar vicinity. There are several studies focused only on the M-dwarf distribution (e.g. \cite{pirzkal09, ryan11, holwerda14}). It is however too early for a clear consensus on how M-dwarfs are distributed to have emerged. This study seeks to constrain the number densities of distant M-dwarfs using the star counts in deep NIR imaging. Furthermore, we apply the results to the extrasolar planet survey data in order to estimate the number of field stars (i.e. non-companions) in a survey field.

In Section~\ref{sec:model}, we explain the employed stellar distribution model of the Galaxy. In Section~\ref{sec:data}, we discuss the used observational data for model optimization. Section~\ref{sec:results} contains a comparison of the observations with the results predicted by the stellar distribution model and a discussion on how we modified the model's parameters to agree with the observational data. Then, this model is applied to the extrasolar planet survey data in Section~\ref{sec:application}.

\section{Model}\label{sec:model}
\subsection{Stellar Distribution Model of the Galaxy}\label{subset:sdm}

We employed the standard stellar distribution model described as follows. There are three components in the model: the thin disk ($D_1$), thick disk ($D_2$), and halo ($H$). In this study, we did not include a bulge component because no observations toward the Galactic center were used (see also Section~\ref{sec:data}). Using the Galactocentric coordinate system, the stellar number density ($n_i$) at position $(R, Z)$ is written as the sum of three components, where $i$ and $n_{\Sol,i}$ represent the spectral type and the corresponding local space density (LSD).
\begin{equation}
n_i (R,Z)=n_{\Sol,i}[D_1(R,Z)+D_2(R,Z)+H(R,Z)]
\end{equation}
An exponential profile and a spheroidal profile are used to describe the distributions of the disks ($D_1$ and $D_2$) and the halo, respectively. The distributions of the number density are written as follows (e.g. \cite{chang11}):
\begin{eqnarray}
&Thin\:Disk&: D_1(R,Z)= (1-f_d-f_h) \times \exp \left[ -\frac{R-R_{\Sol}}{h_{R,1}} - \frac{|Z|-Z_{\Sol}}{h_{Z,1}} \right] \\
&Thick\:Disk&: D_2(R,Z)= f_d \times \exp \left[ -\frac{R-R_{\Sol}}{h_{R,2}} - \frac{|Z|-Z_{\Sol}}{h_{Z,2}} \right] \\
&Halo&: H(R,Z)= f_h \times \left[ \frac{R^2+(Z/\kappa)^2}{R_{\Sol}^2+(Z_{\Sol}/\kappa)^2} \right] ^{-p/2} ,
\end{eqnarray}
where $h_R$ and $h_Z$ represent the scale length and scale height. Subscripts 1 and 2 denote thin disk and thick disk, respectively. $f_d$ and $f_h$ represent the fractions of stars near the Sun that belong to the thick disk and halo component. $\kappa$ is the flattening parameter and $p$ is the power law index of the halo. The position of the Sun is given by $(R_{\Sol}, Z_{\Sol})$.

The number of stars ($N_i$) is calculated for each sub-spectral type using the above model. We define $a$ as the distance from the Sun to position $(R, Z)$. Introducing the Galactic coordinates of the observed field $(l, b)$, the position can be written as follows:
\begin{eqnarray}
R &=& \left[ R_{\Sol}^2+a^2\cos^2b-2R_{\Sol} a\cos b\cos l \right] ^{1/2} \\
Z &=& a\sin b+Z_{\Sol} .
\end{eqnarray}
We estimated the number of stars whose apparent magnitude is between $m_1$ and $m_2$. For each spectral type, apparent magnitudes ($m_1$ and $m_2$) correspond to distances ($a_1$ and $a_2$). The number of these stars in a field ($S$) is calculated by
\begin{equation}
N_i = S \times \int_{a_1}^{a_2} n_i(a)a^2 \:da .
\end{equation}
Note that this integral only depends on the distance $a$. The total number of predicted stars ($N$) is calculated by summation over the number of stars of all spectral types:
\begin{equation}
N = \sum_i N_i .
\end{equation}

\subsection{Parameters of the Stellar Distribution Model} \label{subsec:parameters}

Many studies were conducted to determine the parameters of the stellar distribution model including its scale heights, fractions, and LSDs. \citet{chang12} estimated parameters based on the 2MASS catalogue. We believe these parameters to be the most reliable among those currently available. Fractions of respective components ($f_d$ and $f_h$) are the ratios as compared to the local stellar density. However, because in \citet{chang12} the corresponding values are compared to the thin disk density, we needed to convert these values, and the employed parameters are shown in Table~\ref{tbl-1}. All stars are assumed to follow the same distribution, regardless of their spectral types. The LSDs are provided in the next section.

\begin{table}
  \caption{Parameters of the stellar distribution model.}\label{tbl-1}
  \begin{center}
    \begin{tabular}{ccc}
      \hline
      Component & Symbol & Value \\
      \hline
      Position of the Sun & $R_{\Sol}$ & 8000 pc \\
      & $Z_{\Sol}$ & 25 pc \\
      Thin Disk & $h_{R,1}$ & 3700 pc \\
      & $h_{Z,1}$ & 260 pc \\
      Thick Disk & $h_{R,2}$ & 5000 pc \\
      & $h_{R,2}$ & 1040 pc \\
      & $f_{d}$ & 0.091 \\
      Halo & $\kappa$ & 0.55 \\
      & $p$ & 2.6 \\
      & $f_{h}$ & 0.002 \\
      \hline
    \end{tabular}
  \end{center}
\end{table}

\subsubsection{Local Space Density (LSD)}\label{subsec:lsd}

\begin{figure}
 \begin{center}
 \includegraphics[clip,width=12cm]{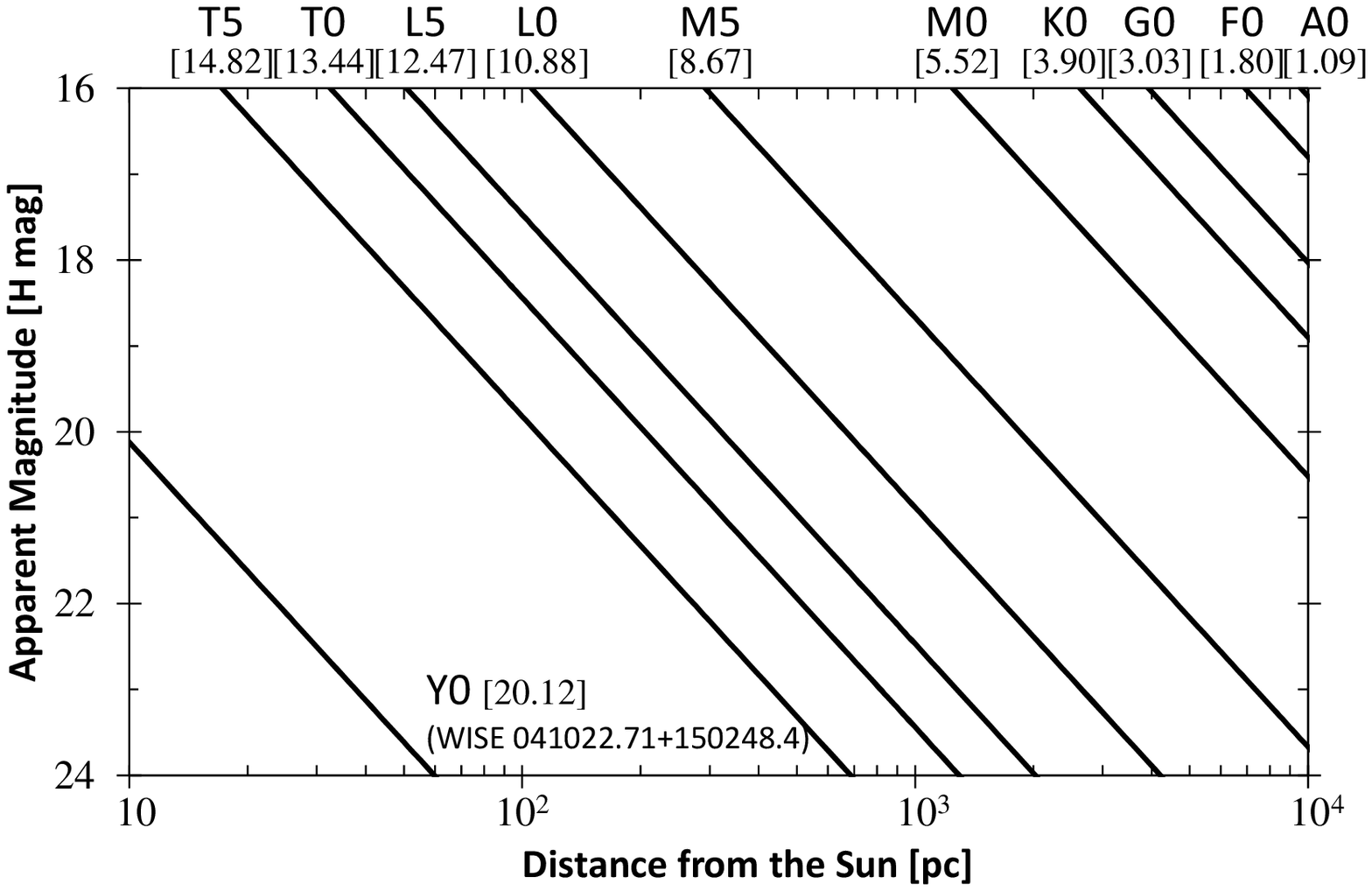}
 \end{center}
 \caption{Relationship between the distance from the Sun and the apparent magnitude. Each line shows a representative spectral type in this magnitude range. The absolute magnitude of a Y0-type star was taken from \citet{kirkpatrick12}. Each value in parentheses represents the absolute magnitude in the $H$ band. Giants are out of range. It is evident from this figure that contamination by giant stars is negligible in the observed magnitude range. \label{fig-1}}
\end{figure}

Our data only includes targets fainter than 16~mag (see also Section~\ref{sec:data}). The focus of our discussion is on the region of $H>20$~mag. As Figure~\ref{fig-1} shows, stars in our sample with magnitudes fainter than 20~mag are mainly M-dwarfs and brown dwarfs. We therefore confine the spectral type of our sample stars to the range from K2 to T7. The local number density of G-dwarfs is $\sim27\%$ compared to that of K-dwarfs (calculated from values in \cite{kirkpatrick12}). Taking the distance and stellar distribution functions into account, the number of G-dwarfs in the 16.5--17.5 bin is estimated to be $\sim35\%$ compared to K-dwarfs, which means that G-dwarfs occupy $\sim3.5\%$ of that bin. F-dwarfs (and earlier type stars) constitute an even smaller fraction. The number of giants is also negligible because giants need to be distant to become faint enough to enter the observed magnitude range (out of range in Figure~\ref{fig-1}). Such giants are very rare.

\begin{figure}
 \begin{center}
 \includegraphics[clip,width=12cm]{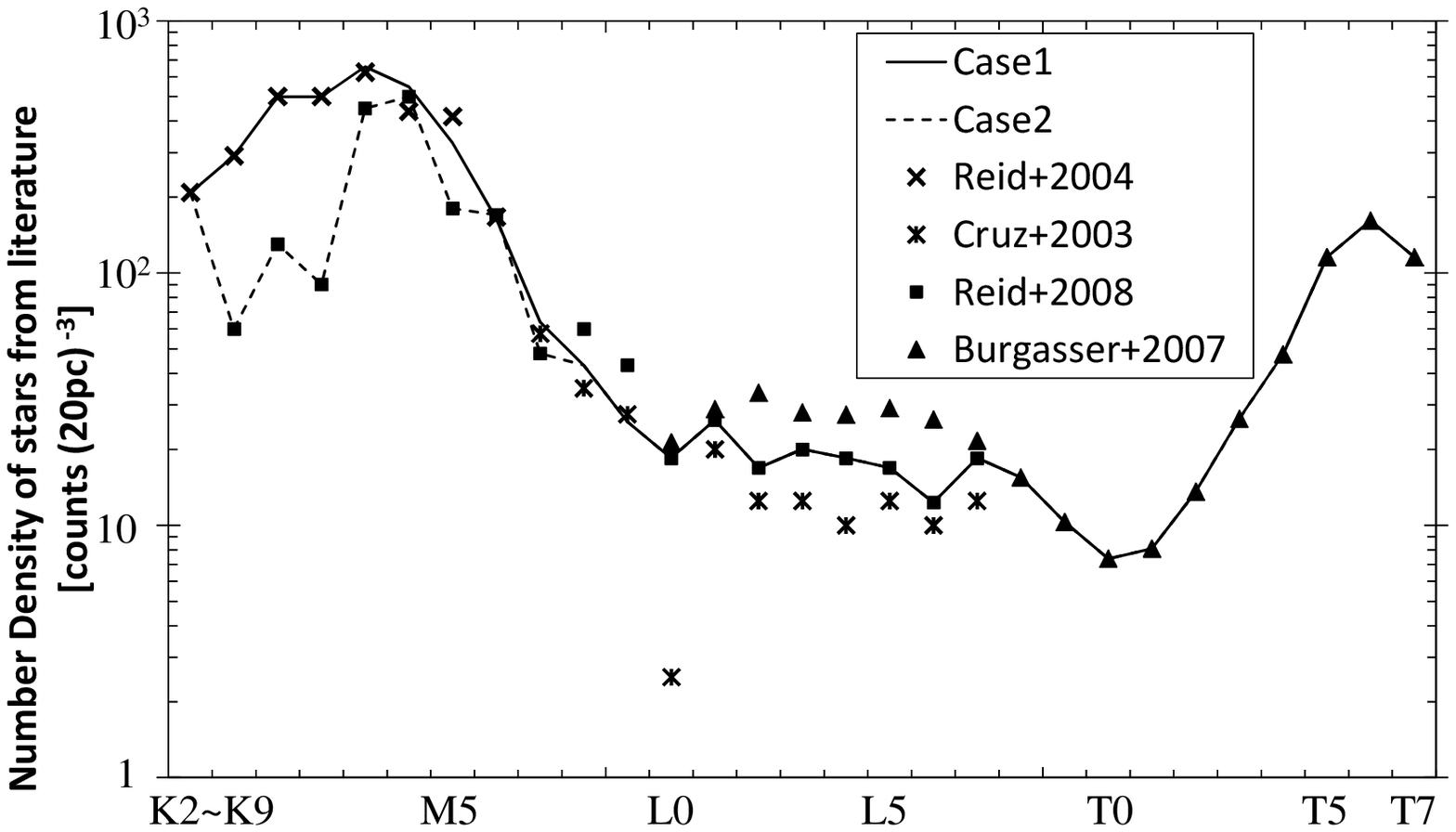}
 \end{center}
 \caption{The number of stars taken from previous studies. The number is uniformly scaled to $(20\:pc)^3$. The dashed line and the dotted line represent Case~1 (standard LSD) and~Case 2 (minimum LSD), respectively. In the range from L0 to L7, we use recent observational results, instead of the values predicted by the model. \label{fig-2}}
\end{figure}

\begin{figure}
 \begin{center}
 \includegraphics[clip,width=12cm]{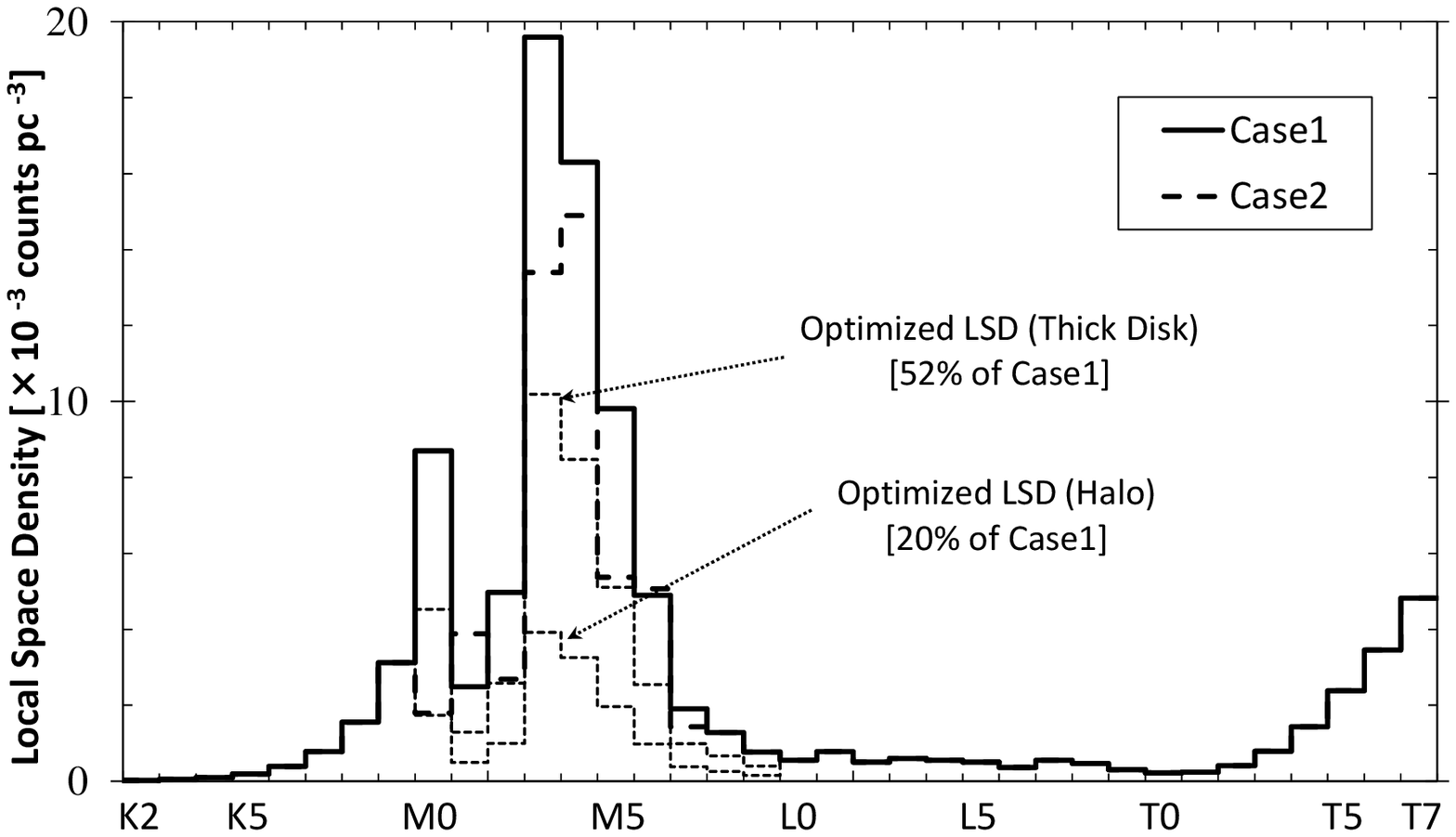}
 \end{center}
 \caption{Local space density of each spectral type. The solid line and the dashed line represent Case~1 (standard LSD) and Case 2 (minimum LSD), respectively. Dotted lines show the optimized LSD, which is discussed in Section~\ref{sec:results}. \label{fig-3}}
\end{figure}

We referred to five studies, described below, to establish the LSD for each sub-spectral type. The number of local stars published in these studies is shown in Figure~\ref{fig-2}. Two cases were adopted; Case~1 is the standard LSD, and Case~2 is the minimum LSD. These cases are shown in Figure~\ref{fig-3}.

\begin{enumerate}
\renewcommand{\theenumi}{(\roman{enumi})}
\renewcommand{\labelenumi}{\theenumi}
\item{{\it \citet{reid04}:}\label{lsd1}} We calculated the LSD from K2 to M6 using this study's luminosity function in the $J$ band. \citet{reid04} mentioned that $J$-band magnitudes are good indicators of the spectral types. The number of each spectral type was estimated using this relation. The spectral type bins used by \citet{reid04} were in some cases broader than the width used in this and other works (e.g. M1, M2 are one bin). We split them into sub-bins in order to achieve equidistant spacing between all spectral bins. The numbers were allocated to the sub-bins in the ratio of 1:2 in favor of fainter stars in cases the exact ratio was unknown. One consequence of this approach is that the LSD did not connect smoothly along the boundaries between M0--M1 and M2--M3. However, this assumption had very little effect on our results, especially in the case of K-dwarfs.
\item{{\it \citet{cruz03}:}\label{lsd2}} This survey examined the number of stars from M7 to L8. Few stars with spectral types later than L0 were included in their analysis because the survey was incomplete at such faint magnitudes. We focused on M7--M9 dwarfs for which we calculated the LSD considering the survey's completeness and sky coverage.
\item{{\it \citet{reid08}:}\label{lsd3}} The number of K through L7 stars was examined in this paper. The focus was mainly on types raging between M9 and L7, while the other sub-spectral types were adopted from the previous work \citep{reid04}. The LSD was calculated using the procedure in \ref{lsd2}.
\item{{\it \citet{burgasser07}:}\label{lsd4}} Stars later than L8 were rarely detected because of their faintness. Theoretical values were therefore used. \citet{burgasser07} made theoretical model calculations of the binary fraction of L- and T-type stars and thereby estimated their number densities. We used the number densities for a primary star with a mass function index of $-0.5$, a mass ratio index of 4, and a binary fraction of 0.1.
\item{\it \citet{caballero08}:} \citet{caballero08} estimated the contamination by field late M-dwarfs and brown dwarfs (L- and T-type stars) in deep surveys. For this they calculated the LSD of M3--T7 quoted from \ref{lsd1}, \ref{lsd2}, \ref{lsd3}, \ref{lsd4}, and \citet{kirkpatrick94}. Their LSD was compatible with our standard LSD (Case~1).
\end{enumerate}

Our LSDs are consistent with other recent studies investigating stellar counts of faint stars (e.g. \cite{ryan11, pirzkal09}), who also employed the above mentioned works. The derived LSD also matches with \citet{kirkpatrick11} for stars later than T6. On the other hands, LSDs later than L5 established in \citet{reyle10} are slightly smaller than ours. We will discuss how  this affects our results in Section~\ref{sec:results}. \citet{holwerda14} suggested much larger central densities for M-dwarfs. However, as these densities are in conflict with the majority of other works we only note them for completeness.

\subsection{Relationship between Spectral Type and Absolute Magnitude}

\begin{figure}
 \begin{center}
  \includegraphics[clip,width=12cm]{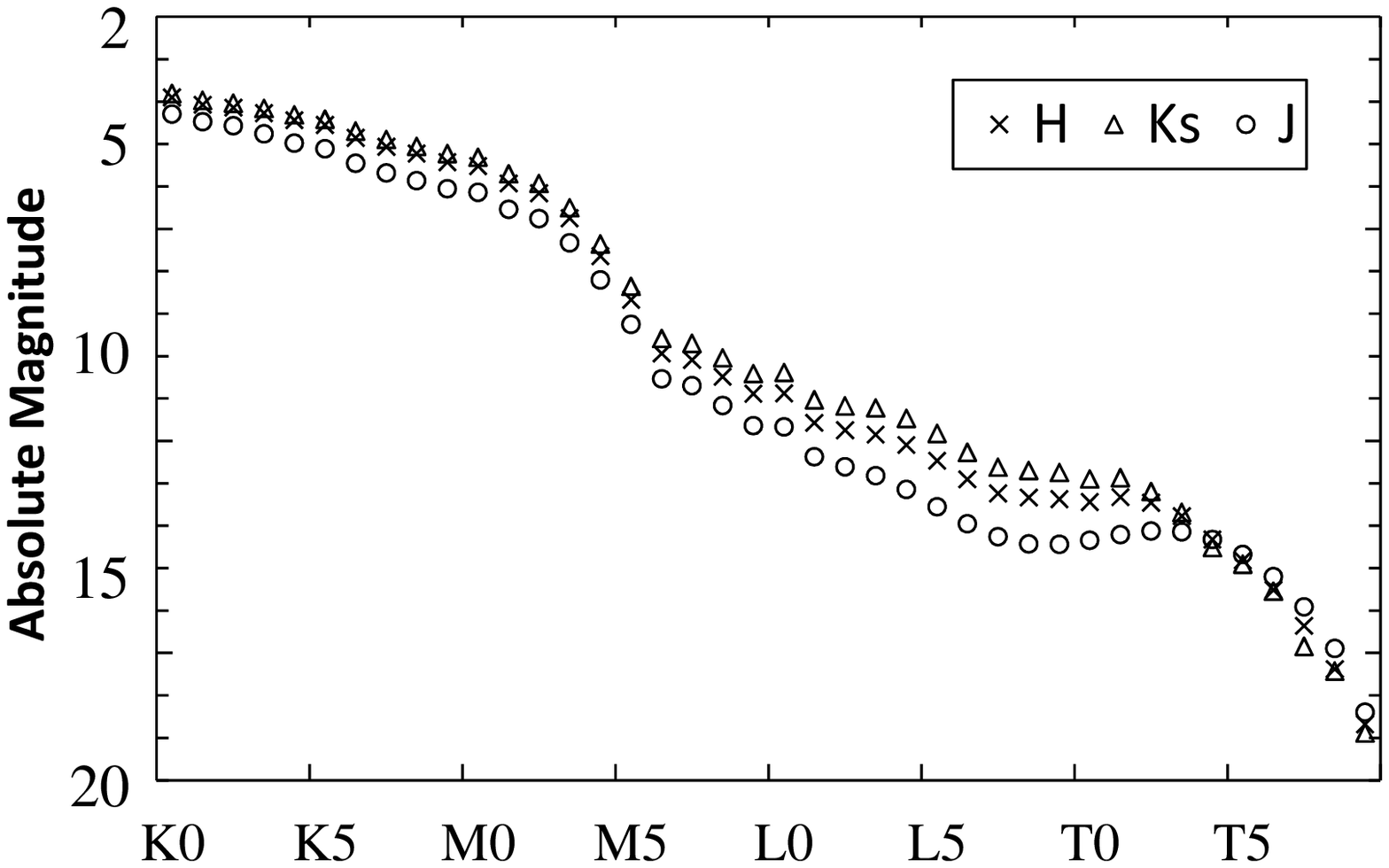}
 \end{center}
 \caption{Relationship between spectral types and absolute magnitudes in the $H$, $K_s$, and $J$ band taken from Mamajek's list (see URL of footnote~\ref{f_mamajek}). \label{fig-4}}
\end{figure}

The typical magnitudes of each spectral type were quoted from the list\footnote{http://www.pas.rochester.edu/\~{}emamajek/EEM\_dwarf\_UBVIJHK\_colors\_Teff.txt\label{f_mamajek}} provided by E. Mamajek as at June 18th 2013. This list compiles information about stellar characteristics, such as colors, magnitudes, and effective temperatures, on the basis of spectral standards and other studies (see also \cite{pecaut12} and \cite{pecaut13} for the stellar information). Figure~\ref{fig-4} shows the relationship between spectral type and absolute magnitude in the $H$, $K_s$, and $J$ band. We focus our discussion on the $H$-band results. The $K_s$- and $J$-band magnitudes are used only to confirm the $H$-band results (see Section~\ref{subsec:results_goods-n} and Section~\ref{subsec:results_goods-s}).

\section{Data}\label{sec:data}

The magnitude of planet candidates is generally $H<22$~mag. In order to evaluate the aforementioned model for such faint star counts, we used archived data from deep and wide $H$-band imaging surveys: the GOODS (Great Observatories Origins Deep Survey) North and South fields. In this section, we explain how the data was gathered and analyzed.

\subsection{The GOODS-North Field}

We referred to the MODS (MOIRCS Deep Survey) region of the GOODS-North field ($l=125.9\arcdeg$, $b=+54.8\arcdeg$), observed by \citet{kajisawa11} in the $J$, $H$, and $K_s$ bands with Subaru's Multi-Object Infrared Camera and Spectrograph (MOIRCS: \cite{ichikawa06, suzuki08}). The FoV of MOIRCS is $4\arcmin \times 7\arcmin$ and four pointings covered an area of $\sim105$ square arcminutes in the survey. Only the detected number in the $H$ band was compared with the predicted number. Stars were selected from this catalogue based on the method used in \citet{kajisawa11}. This catalogue was released on the MODS homepage\footnote{http://www.astr.tohoku.ac.jp/MODS/wiki/}.

\citet{kajisawa11} performed photometry using different aperture sizes depending on whether they measure colors (aperture magnitudes) or the total flux (total magnitudes). We only used $H$-band results. However, the total magnitudes were given only in the $K_s$ band, and only the aperture magnitudes were available for the $H$ band. We therefore performed the aperture correction for the $H$-band magnitudes using the same difference between the total and aperture magnitudes as in the $K_s$ band. The distinction between stars and galaxies was done based on the spectroscopic work by \citet{wirth04}. It is difficult to distinguish between L- and T-type brown dwarfs and galaxies by spectroscopy, but the result should remain largely unaffected because the total number of such stars detected in this survey was expected to be small. We only used stars brighter than 21.5~mag because colors and spectroscopic observations were not sufficient for fainter targets. We thus assumed all objects brighter than 20.5~mag to be stars, although some contamination (e.g. external galaxies) is to be expected in the 20.5 to 21.5~mag bin. Figure~\ref{fig-5-6}(a) shows the comparison between the detected number and the predicted number per square degree. The uncertainty is given as $1\sigma$ Poisson scatter.

\subsection{The GOODS-South Field}

We used two regions of the GOODS-South field ($l=223.6\arcdeg$, $b=-54.4\arcdeg$): the ERS (Early Release Science) region \citep{windhorst11} and the CANDELS (Cosmic Assembly Near-infrared Deep Extragalactic Legacy Survey) region \citep{guo13}. The ERS region was observed by \citet{windhorst11} using the Wide Field Camera 3 aboard the Hubble Space Telescope. Only data taken in the F160W filter (corresponding to the $H$ band) was used. The total sky coverage was $9\farcm75 \times 4\farcm5$. Stars have already been distinguished from galaxies by \citet{windhorst11}. The CANDELS region was observed using the same instrument, achieving a total area of $\sim165$ square arcminutes. We extracted stars from the point source catalogue created by \citet{guo13} based on the value of {\it CLASS\_STAR}. Objects whose {\it CLASS\_STAR} was larger than 0.8 were regarded as stars. We only used the photometric results of the F160W filter data ({\it FLUX\_AUTO}). We converted the flux in the $AB$ magnitude system to the Vega magnitude system (adopting a correction value of 1.5~mag only for the ERS region). The final detection limit was approximately 24.5~mag in the ERS region, and 23.5~mag in the CANDELS region. Therefore, we assumed that all objects in the ERS region and all, except those in the 22.5--23.5~mag bin within the CANDELS region are stars, and expected that the 22.5--23.5 mag bin in the CANDELS region includes some external galaxies. Figure~\ref{fig-5-6}(a) shows the comparison between the detected number and the predicted number per square degree. While the error bars of the predicted number represent $1\sigma$ Poisson scatter, the uncertainties in the detected number of the ERS region are taken from \citet{windhorst11}.

\section{Results and Discussion}\label{sec:results}

\begin{figure*}
 \begin{center}
  \includegraphics[clip,width=18cm]{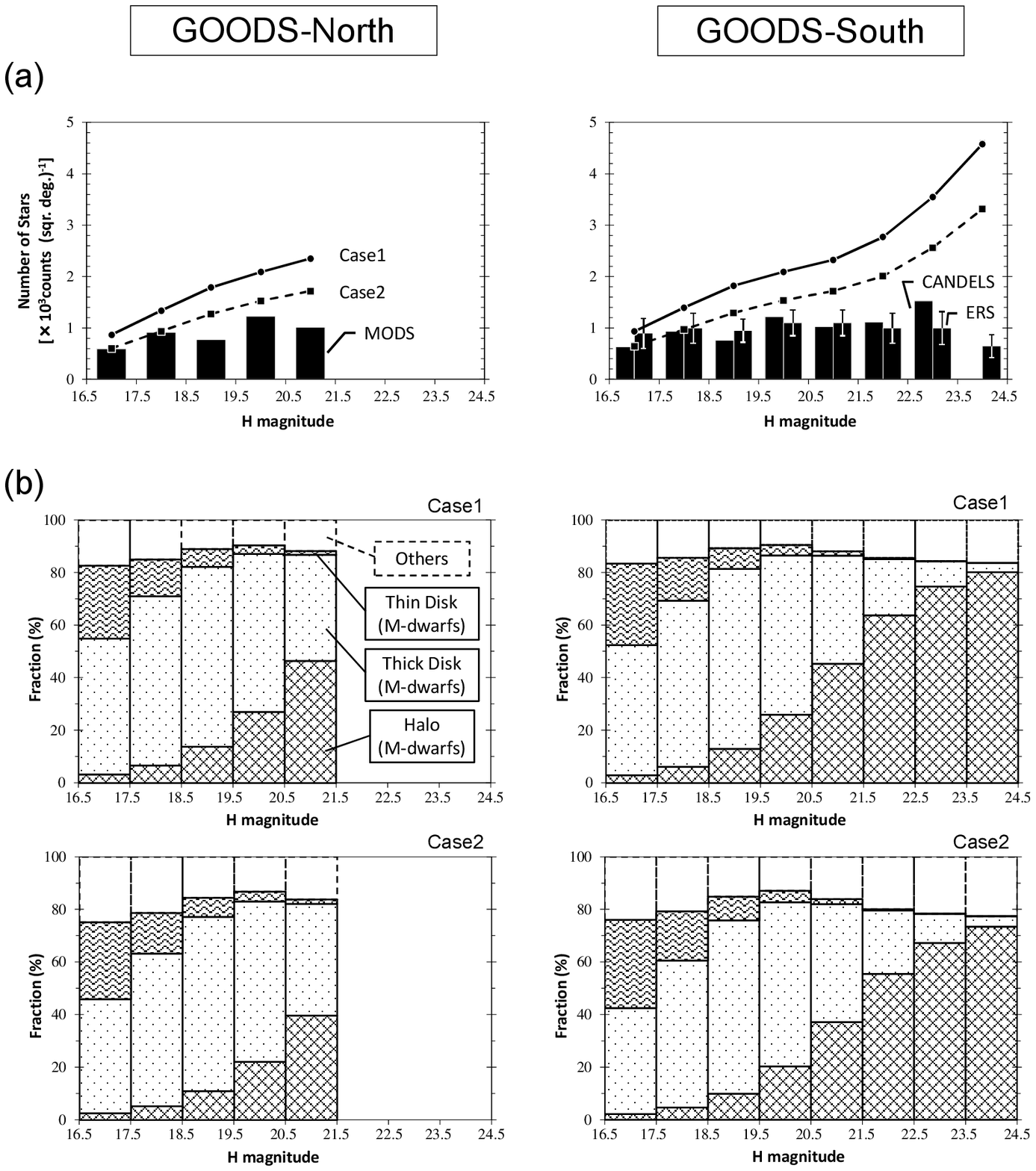}
 \end{center}
 \caption{(a) Comparison between the number of detected and predicted stars. The bar graph shows the detected number. The solid and dashed lines show the prediction from Case~1 and Case~2, respectively. Symbols of Case~1 and Case~2 hide each error bar. (b) Fraction of Galactic components (thin disk, thick disk, and halo) relative to all predicted numbers. The different types of gray/shaded regions show the fraction of M-dwarfs in the Galactic components, whereas the unshaded regions contain all the other spectral types. \label{fig-5-6}}
\end{figure*}

The results of GOODS-North (MODS region) and GOODS-South (ERS and CANDELS regions) are shown in the left and right side of Figure~\ref{fig-5-6}, respectively. In Figure~\ref{fig-5-6}(a), the detected numbers shown as bar graphs were compared with predicted numbers shown as a solid line (Case~1: standard case) and dotted line (Case~2: minimum case). Figure~\ref{fig-5-6}(b) shows fractions of Galactic components (the thin disk, thick disk, and halo) of M-dwarfs and other spectral types relative to all predicted numbers. There is a slight difference between the predicted numbers for Case~1 and Case~2 in Figure~\ref{fig-5-6}(a), but the features for these two cases resembled each other when compared with observations. We therefore only discuss the results with respect to the predictions of Case~1, while the same conclusion holds for Case~2. It was not necessary to correct for interstellar extinction because observations were conducted in the $H$ band and the Galactic pole data was used.

We mention the results of comparison of each field in Subsection~\ref{subsec:results_goods-n} and \ref{subsec:results_goods-s}, then consider how to optimize the stelar distribution model to match with these deep NIR imaging in Subsection~\ref{subsec:discussion}.

\subsection{Results of the GOODS-North Field}\label{subsec:results_goods-n}

\begin{figure*}
 \begin{center}
  \includegraphics[clip,angle=270,width=18cm]{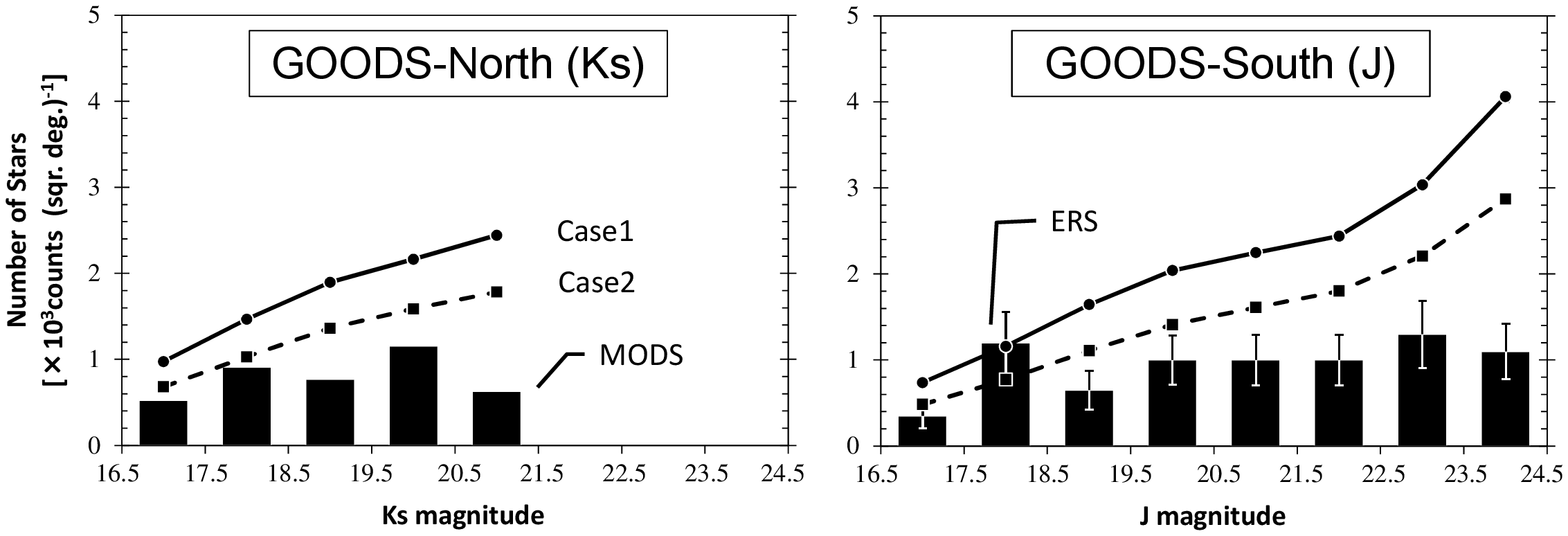}
 \end{center}
 \caption{Comparison between the number of detected and predicted stars. The left panel shows the $K_s$-band result of the MODS region and the right shows the $J$-band result of the ERS region. Legends and the field size are the same as in Figure~\ref{fig-5-6}(a). \label{fig_new}}
\end{figure*}

The detected number was smaller than the predicted number in all bins in the left panel of Figure~\ref{fig-5-6}(a). We found from Figure~\ref{fig-5-6}(b) that almost all predicted stars are members of the thick disk or the halo, and that these stars are mainly M-dwarfs. This shows that the number of M-dwarfs in the thick disk and halo component is small compared to the model predictions. This result indicates that both the thick disk and halo have fewer M-dwarfs compared to the population in the solar vicinity. We also checked the $K_s$-band star count using the LSD shown in Figure~\ref{fig-3} and the relationship shown in Figure~\ref{fig-4}. The left side of Figure~\ref{fig_new} shows the comparison between the detection and prediction. The same result holds if MODS data from the $K_s$ band is used.

Stars other than M-dwarfs (e.g. L- and T-type stars) were rarely detected in the range of 16.5 to 21.5~mag (see Figure~\ref{fig-5-6}(b)). Moreover, contamination from external galaxies is negligible because \citet{wirth04} distinguished stars from galaxies based on their spectra. While the possibility of contamination by faint galaxies cannot be completely ruled out at magnitudes between 20.5 and 21.5~mag, our conclusion is conservative because the difference between the detected and predicted number would be even larger than that in case of including external galaxies by mistake, further enhancing the problem of missing faint stars in the thick disk and halo.

\subsection{Results of the GOODS-South Field}\label{subsec:results_goods-s}

The detected number was consistent with the predicted number for magnitudes brighter than 17.5~mag, but much smaller for fainter magnitudes in both ERS and CANDELS regions from Figure~\ref{fig-5-6}(a). Almost all predicted stars belong to the thick disk or halo, and most of these stars are M-dwarfs according to the model (see Figure~\ref{fig-5-6}(b)). This finding implies that fewer M-dwarfs exist in the thick disk and the halo compared to the prediction using the standard LSD. This tendency is corroborated by observations from the GOODS-South field using the F125W filter (corresponding to the $J$ band), as can be seen in the right panel of Figure~\ref{fig_new}. The finding is thus independent of the observational band.

The catalogue by \citet{windhorst11} distinguished between stars and galaxies reliably. All detected objects shown in the ERS region of Figure~\ref{fig-5-6}(a) are therefore regarded as stars. In the CANDELS region, stars and galaxies were identified based on the $CLASS\_STAR$ values. It is generally difficult to avoid contamination by galaxies completely for faint objects when this method is employed. However, our conclusion as mentioned above does not change because the discrepancy between the detected and predicted stellar counts would only be enhanced, even in case of a small fraction of extragalactic contaminations. There was a discrepancy between the ERS and CANDELS star counts. Focusing on the 22.5--23.5 mag bin, the star count of CANDELS was larger than that of ERS. We believe this to be due to the inclusion of some galaxies in the CANDELS star counts. All other bins were consistent within the ERS uncertainties.

\subsection{Discussion}\label{subsec:discussion}

The detected number of Galactic stars is significantly smaller than the number predicted by the standard stellar distribution model of the Galaxy. Most of the stars belong to the thick disk or halo components and more than 80\% of these stars are M-dwarfs according to the model. The detected stellar count and the predicted numbers from a standard parameterization of the Galaxy begin to differ significantly at $H\sim20.5$~mag, a depth previously unachieved by surveys traditionally used for constraining the Galactic parameters (SDSS, 2MASS). Detected M-dwarfs fainter than 20.5~mag need to be more distant than $\sim1$~kpc from the Sun. Our results imply that there are significantly fewer M-dwarfs at 1~kpc or more from the Galactic plane than previously thought. The number density of M-dwarfs in the thick disk or the scale height of the thick disk has to be decreased to match the model's results with the detected number in both GOODS fields. In addition to this, the number density of M-dwarfs in the halo has to be decreased. Alternatively, employing significantly smaller fractions of the halo ($f_h$) and thick disk ($f_d$) can also explain the scarcity of M-dwarfs observed. As the deep NIR data we used is predicted to contain mostly distant M-dwarfs (see Figure~\ref{fig-5-6} b), both choices have similar effects in our model. \citet{hayden13} showed, for stars up to 3~kpc, that the stellar metallicity decreases with increasing distance from the Galactic mid-plane, which is explained by the scenario of halo and thick disk originating from dwarf galaxies having merged with the Galaxy (e.g. \cite{searle78, abadi03, kirby08}). It is theoretically predicted that the cooling rate is small in metal-poor clouds and that high-mass stars tend to form more than low-mass stars in this condition. Stars belonging to the halo may form from metal-poor clouds. With respect to our results, this indicates that lower metallicity in the halo and thick disk component might be correlated with a M-dwarf deficit. 

\begin{figure*}
 \begin{center}
   \includegraphics[clip,angle=270,width=18cm]{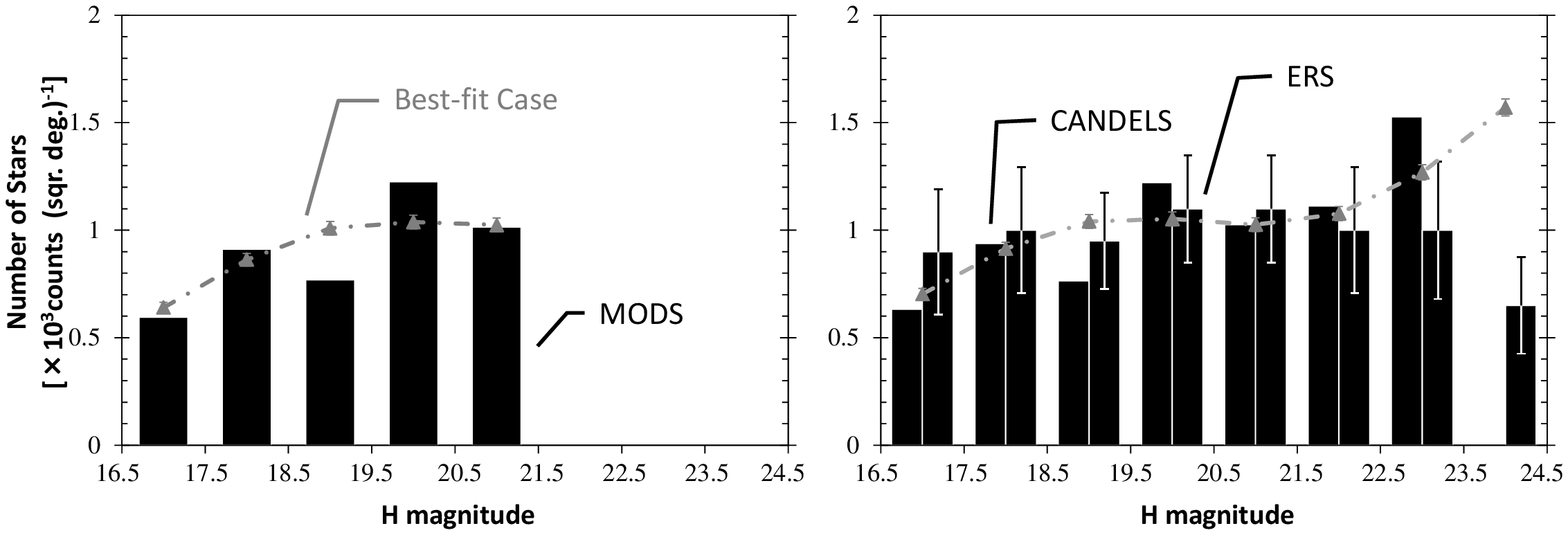}
  \end{center}
  \caption{Observed and predicted numbers of stars, using the best-fit number densities of M-dwarfs in the halo and the thick disk (see Subsection~\ref{subsec:discussion}). The gray dot-dashed line represents the predicted number. Other legends and the field size are the same as in Figure~\ref{fig-5-6}(a). \label{fig-7}}
 \end{figure*}

We used three methods to improve upon the stellar distribution model's parameters. Firstly, we changed the number densities of M-dwarfs (M0--M9) in the halo and thick disk, in order to match them with the detected number. Only the ERS region of the GOODS-South data was used because of its small uncertainty and large sample size. The best-fit case (shown in Figure~\ref{fig-3} as dotted lines) was achieved with number densities of $20 \pm 13\%$ and $52 \pm 13\%$ for the halo and the thick disk, respectively, relative to the standard parameters (Case~1). We re-calculated the predicted numbers using the best-fit case, and show the comparison between the new predictions and the detected numbers in Figure~\ref{fig-7}. It is evident that the best-fit case in Figure~\ref{fig-7} improves on the standard parameters shown in Figure~\ref{fig-5-6}(a). In Figure~\ref{fig-7}, a small disagreement can be seen in some bins, but this discrepancy falls within the $3\sigma$ uncertainty, considering Poisson noise of predictions and observational uncertainties. As mentioned in subsection~\ref{subsec:lsd}, our LSDs for stars later than L5 are slightly larger than those of \citet{reyle10}, but the overall effect on our results is small considering the expected low number of these stars. 
Results of \citet{pirzkal05} also tentatively pointed into the direction of a deficit in the late-type halo subdwarf population. They showed that the late M-dwarf population could not be reproduced with their assumed halo to disk ratio (0.25\%) and axial ratio of the halo (0.7). While we did not discuss the bulge, \citet{sumi11} showed that the number density of M-dwarfs in the bulge is largely similar to the mass function from \citet{chabriere03}.

Secondly, we varied the halo and thick disk fractions ($f_h$ and $f_d$). When smaller fractions were employed, the number of predicted stars was likewise reduced. Considering that most stars of 16.5--24.5 mag were M-dwarfs, employing overall smaller halo and thick disk fractions gave almost the same result as when reducing the number density of M-dwarfs in the halo and thick disk. However, even if we use the minimum fractions for the halo and thick disk within the uncertainty given by \citet{chang11}, the discrepancy between the predicted and detected number still emerges. We still need to reduce the number density of M-dwarfs to $58\%$ and $64\%$ of the original value for the thick disk and halo to be within the bound of uncertainty given by \citet{chang11}.

Finally, we changed the scale height and length of the thick disk using the standard number density (Case~1). The scale height of the thick disk was reported in many studies (including \cite{chang11, polido13, carollo10} and the references therein), based on star counts and stellar kinematics of the Galaxy. The scale height varies between 500 and 1600~pc in these studies. \citet{carollo10} showed that the scale height of the thick disk is $990 \pm 40$~pc using stellar kinematics of the Galaxy. Most of the previous studies based on star counts (see references in \cite{chang11}) concluded that the typical scale height is about $\sim1000$~pc. Some studies, however, determined a smaller scale height. \citet{polido13} for example stated that the thick disk's scale height is $640 \pm 70$~pc based on 2MASS star counts in the $J$, $H$, and $K_s$ bands. We found that the predicted number was consistent with the detected number when the disk parameters from \citet{polido13} were employed. In addition, several works focused only on the late-type dwarf distribution (e.g. \cite{pirzkal09, ryan11, holwerda14}) estimated the scale height of the thin disk based on star counts using a single disk model. They all concluded that the scale height of early M-dwarfs is larger than that of later than M5-dwarfs. These results indicate that the thick disk component of late-type dwarfs cannot be clearly distinguished from the thin disk component. Similarly, \citet{bovy12} concluded that there is no thick disk in the Galaxy, which is equivalent to employing a small scale height of the thick disk. Moreover, there is some variation in the scale length of the disks in the published literature (e.g. \cite{bensby11}). We changed the scale length of the thick disk ranging from 2000 to 5000~pc, however, the predicted number was not considerably affected. 

In summary, it can be said that the predicted star count at $H>20.5$~mag was not consistent with standard parameters from models construed from data with relatively low limiting magnitudes (e.g. \cite{chang12}). They tend to over predict the number of distant faint stars. The model matches the observations when we reduce the M-dwarf number density for the halo and thick disk component by $20\pm13\%$ and $52\pm13\%$, respectively. Alternatively, the overall thick disk fraction ($f_d$) and halo fraction ($f_h$) can be lowered. Reducing the thick disk's scale height to $\sim600$~pc also made the model agree better with the data, but still required changing the M-dwarf density (or overall fraction $f_h$) of the halo. The question whether the fraction of M-dwarfs in the thick disk is smaller to begin with, or whether its scale height is lower for late-types can ultimately not be resolved from the presented data.

\section{Application to the Extrasolar Planet Survey Data}\label{sec:application}

In Section~\ref{sec:results}, the model parameters were fit to the deep NIR survey result. We calculated the predicted number of stars, that are not companions, in the fields of the extrasolar planet survey, the SEEDS (Strategic Exploration of Exoplanets and Disks with Subaru) project, by using the best-fit number density.

\subsection{The SEEDS Data} \label{subsec:seeds-data}
\begin{figure*}
 \begin{center}
  \includegraphics[clip,width=18cm]{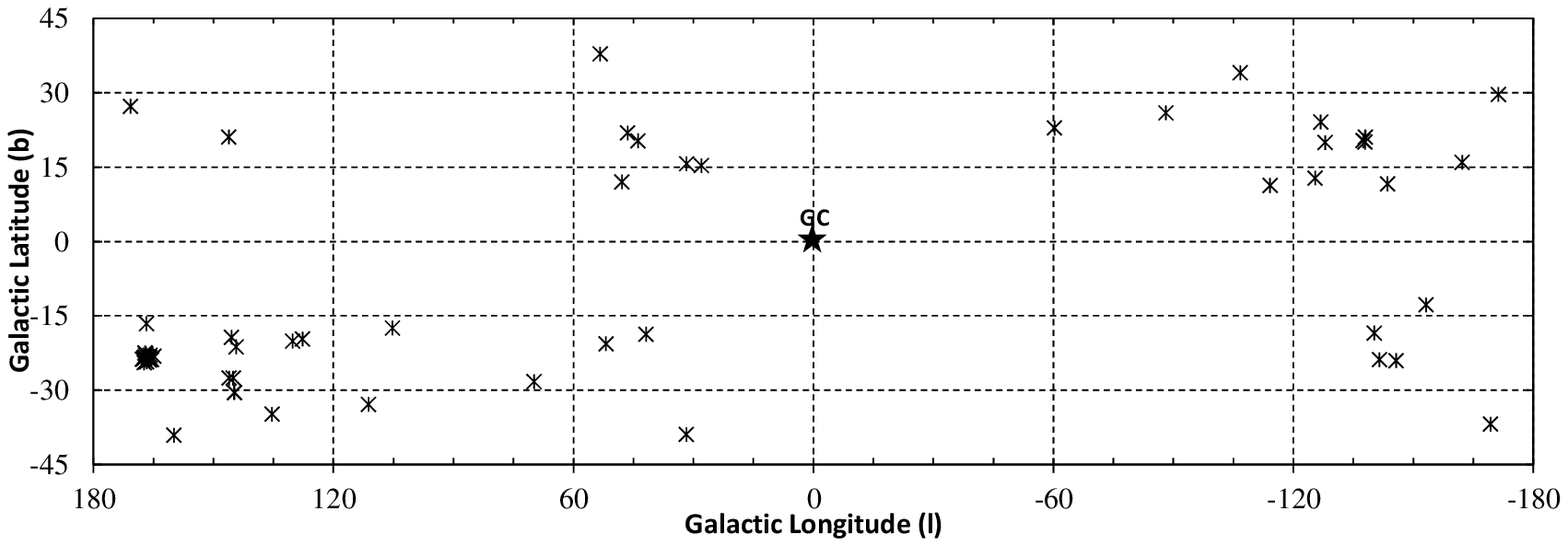}
 \end{center}
 \caption{Galactic coordinates of 62 stars observed in the SEEDS project. We excluded stars near the Galactic plane and/or the Galactic center (GC).\label{fig-8}}
\end{figure*}

In the SEEDS project, extrasolar planets and protoplanetary disks were observed using direct imaging to elucidate universal rules of planetary formation \citep{tamura09}. Observations were conducted using the Subaru telescope together with the high-contrast imaging detector HiCIAO \citep{suzuki10} and the adaptive optics device AO188 \citep{hayano10, hodapp08}. The FoV of a single observation is $19\farcs5 \times 19\farcs5$. From observations conducted between October 2009 and April 2013, 62 stars were selected on the basis of the five criteria listed below. Figure~\ref{fig-8} shows the Galactic coordinates of the selected 62 stars. The total fields measured 6.5 square arcminutes.

\noindent
\begin{enumerate}
\renewcommand{\theenumi}{(\Roman{enumi})}
\renewcommand{\labelenumi}{\theenumi}
\item{{\it Stars with the Galactic coordinates (longitude: $l$, latitude: $b$) in the following range:}\label{seeds1}}
\begin{center}
$20\arcdeg<|l| \:\:\:\cap\:\:\: 10\arcdeg<|b|<40\arcdeg$ .
\end{center}
This requirement excludes stars in the bulge and the Galactic plane subject to heavy extinction. The region of $|b|>40\arcdeg$ was omitted due to small sample size. This omission has little effect on the final results.
\item{\it Stars observed in the $H$ band.} \\ For the SEEDS project, most observations were conducted in the $H$ band, and therefore, only $H$-band data were used.
\item{\it Stars observed using the direct imaging + angular differential imaging (DI+ADI: \cite{marois06}) mode.} \\ The DI+ADI mode is very effective for detecting point sources. We excluded more than 50 stars which were not observed in this mode.
\item{{\it Sufficient exposure time.}\label{seeds4}} \\ The detection limit improves with longer exposure times. We selected only deep observations whose exposure time was longer than 15 minutes.
\item{\it Data is available for analysis.} \\ Despite satisfying requirements \ref{seeds1} through \ref{seeds4}, image data was not available for about 40 targets and could not be included in this study.
\end{enumerate}

\begin{figure}
 \begin{center}
  \includegraphics[clip,width=12cm]{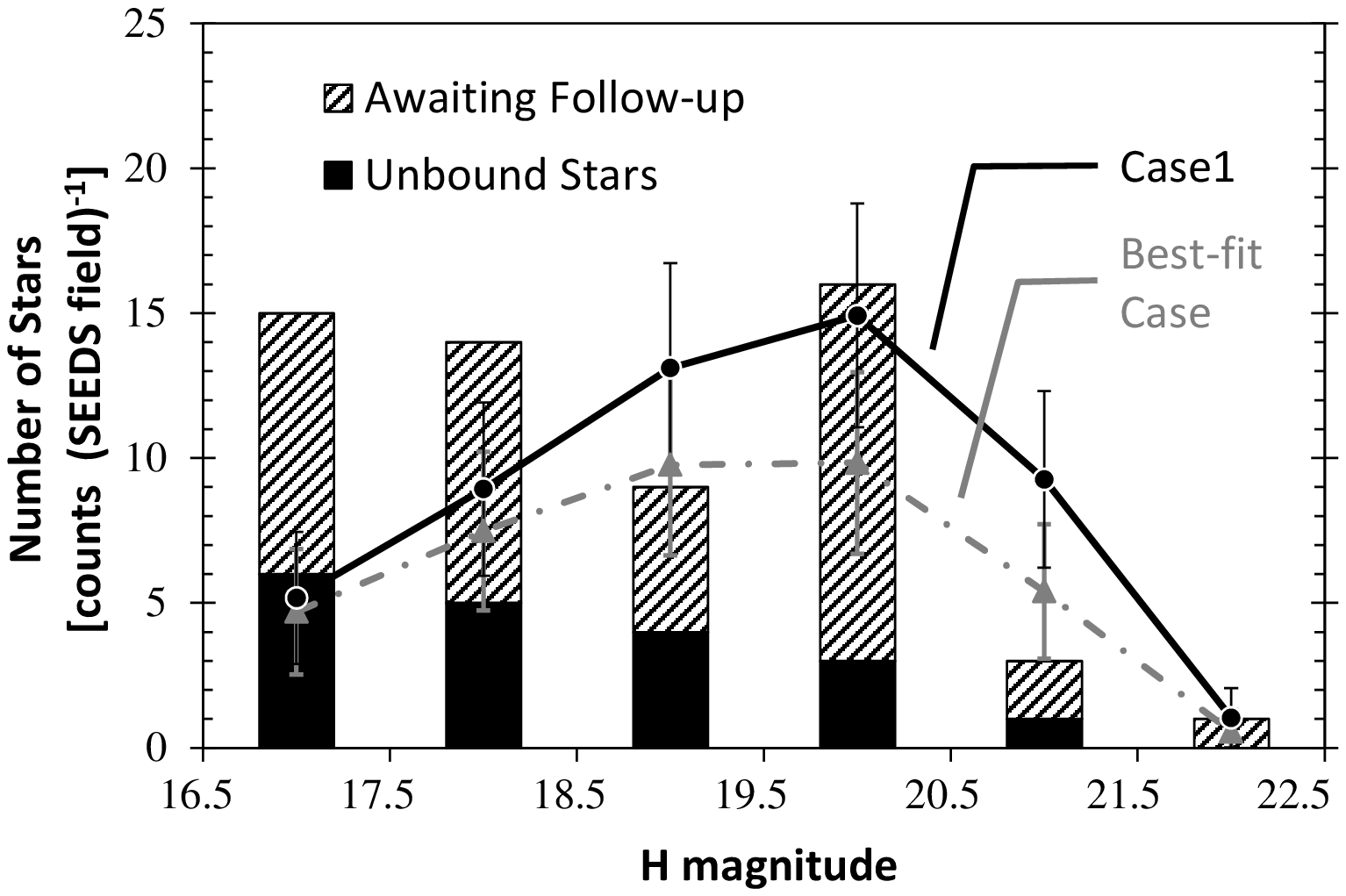}
 \end{center}
 \caption{Observed and predicted number of stars in the SEEDS field, using the best-fit number density (dotted-dashed gray line) and the standard case (Case~1; solid black line). The whole SEEDS field (6.5 square arc minutes) is shown. The detected and predicted number represented by the bar graph and the line graph, respectively. Striped bins show stars whose binary status has not been confirmed yet. \label{fig-9}}
\end{figure}

Point sources detected near primary stars were determined to be companions (binaries or planets) by measuring their proper motions. While it would be preferable to use only confirmed unbound stars, we also included stars whose binary status could not be confirmed to extend the sky coverage and increase the sample size. This is justified by the large separation and faintness of many point sources lacking proper motion confirmation, rendering the a-priori probability of these objects being part of the primary system very small. The analysis method is shown in Appendix~\ref{app:analysis}. We then calculated the number of stars, that are not companions, using the best-fit number density. Figure~\ref{fig-9} shows the comparison between the observed number and the predicted number in the whole SEEDS field, with a given uncertainty of $1\sigma$ Poisson scatter. For reference, we also showed the predicted number using the standard case (Case~1) in Figure~\ref{fig-9}.

\subsection{Discussion on the SEEDS field}
Focusing on the result of the best-fit model, the detected number is consistent with the predicted number at 18.5--19.5~mag and fainter than 20.5~mag when we include the stars awaiting follow up observations (striped bins in Figure~\ref{fig-9}), but is larger in the other bins. The number density in the thin and thick disk may be increased to match with the detected number by being considered that there is little possibility that stars shown as stripe bins are companions, and that there is a few halo stars in the SEEDS field. Increasing the number density in the thin disk leads to a stronger deviation from the result of the GOODS field. The cause of this discrepancy could be explained by considering that the SEEDS field is about ten time smaller than each GOODS field. The model result is still within $3\sigma$ uncertainty of the data. 
We then discuss the difference between the best-fit case and the standard case (Case~1). The predicted number in the standard case is larger for all bins, and might be more consistent with the detected number for magnitudes brighter than 20.5~mag compared to the best-fit case. However, employing the best-fit case is well-suited to reproducing the number of fainter stars ($H>20.5$~mag), because we produced the best-fit case to match with the number of detected faint stars in the GOODS fields. We thus propose employing the model using the best-fit number density, which includes smaller M-dwarf density in the halo and thick disk, when estimating the number of faint contaminating field stars in extrasolar planet surveys.

We further discuss the discrepancy between the detected and the predicted number, especially with respect to stars that have unconfirmed companion status. There is a high probability that most of these stars are unbound stars. \citet{yamamoto13} showed that the fraction of stars with giant planets at larger separation is less than 17.9\%, and an equal or lower fraction was estimated in previous studies (including the references in \cite{yamamoto13}). While we cannot rule out contamination by external galaxies completely due to lack of multi-color observations in the SEEDS project, galaxies can still be excluded based on the FWHM of their PSF (point spread function) in most cases. The true number of detected stars would be smaller than the number shown in Figure~\ref{fig-9}, if any galaxies still remain in the sample of detected stars. 
Other effects, such as interstellar extinction, stars of the bulge component, and contamination by giants, can be neglected due to the selection and the magnitude range of the sample (see Subsection~\ref{subsec:seeds-data} and Figure~\ref{fig-1}). Some stars in the SEEDS field belong to a cluster (e.g. the Pleiades open cluster). Therefore there is a possibility that low-mass stars in the cluster were included. We estimated the number of low-mass stars from the mass function of the Pleiades (e.g. \cite{bihain06}), but the number is too small to explain the disagreement between the detected and predicted numbers. Consequently, it is considered that the discrepancy is attributed to the small sample size because it cannot be fully explained by the aforementioned causes.

\section{Summary}
We inspected predictions from a stellar distribution model from deep NIR imaging, in order to estimate the number of unbound stars in extrasolar planets surveys using direct imaging. We used data from regions in the GOODS-North (MODS region) and GOODS-South fields (ERS and CANDELS regions). The detected number of stars was compared with the predicted number using the standard stellar distribution model in the Galaxy. These numbers begin to differ significantly for $H>20.5$~mag. The number density of M-dwarfs in the thick disk or the scale height of the thick disk have to be decreased to match with the detected number. In addition, M-dwarfs have to be much less abundant in the halo component than they are expected from the number of M-dwarfs in the solar vicinity. Alternatively, significantly reducing the overall halo and thick disk fraction yields similar results. The degeneracies in the model parameters (e.g., reducing scale height or component fraction) cannot ultimately be resolved with deep IR data alone. Regardless of the exact realization chosen, it is clear that previous models based on data not as deep (i.e. SDSS or 2MASS) over predict the number of M-dwarfs in the regions distant from the Galactic plane. This has implications for the prediction of the number of contaminating faint sources in direct imaging surveys for extrasolar planets. We consequently apply the model using our best-fit number density to the data of an actual extrasolar planet survey, the SEEDS data. The detected number of unbound stars shows good agreement with the prediction, but it is larger than expected in two of the magnitude bins (16.5--17.5 and 19.5--20.5). It is thought that the small sample size might cause this discrepancy. We recommend employing the model using the best-fit number density, which has a smaller M-dwarf density in the halo and thick disk, when estimating the number of non-companion faint stars. 


\bigskip

We thank R. Kooistra for spending the time to carefully proof-read our manuscript. We also thank the referee for all the insightful comments that helped improve this paper.

\appendix
\section{Analysis of the SEEDS data} \label{app:analysis}

The samples were analyzed in accordance with the following steps.
\noindent
\begin{enumerate}
\renewcommand{\theenumi}{(\Alph{enumi})}
\renewcommand{\labelenumi}{\theenumi}
\item{{\it ADI analysis:}\label{analysis1}} \\ The purpose of the ADI analysis is to improve the contrast between the primary star and its surroundings by reducing the quasi-static speckle noise resulting from deep exposures. Many methods for the analysis of ADI data have been proposed. In this study, we only considered the region farther than $1\farcs5$ away from the primary star to neglect the influence of the primary star's halo. Standard ADI analysis was employed as it provides the best signal-to-noise ratio in the region farther from the primary star \citep{konishi13}. LOCI (Locally Optimized Combination of Imaging; \cite{lafreniere07}) analysis was not used because this method is often used to probe the closest vicinity of stars. The analysis was implemented in IRAF (Image Reduction and Analysis Facility).
\item{{\it Photometry of the primary star:}\label{analysis2}} \\ The star's magnitudes and the detection limits were calculated using photometry of the primary star. In cases where images of the primary star were saturated, we used the unsaturated data that was taken with the shorter exposure times and with a neutral density filter before and after the saturated images. Aperture photometry was performed with the APPHOT package of IRAF. The FWHM size of the stars was at most 8 pixels. Both 20 and 40 pixels aperture radii were employed and for the sky area we adopted an annulus from 40 to 100 pixels centered on the star. The magnitude of the primary star was taken from the SIMBAD catalogue\footnote{http://simbad.u-strasbg.fr/simbad/sim-fbasic}.
\item{{\it Evaluation of self-subtraction:}\label{analysis3}} \\ The flux of point sources is reduced by self-subtraction in ADI analysis. Self-subtraction generally increases when point sources are near the primary star. However, in this study, self-subtraction is approximately constant because only the area farther than $1\farcs5$ from the primary star was used. We estimated the amount of self-subtraction using the following method: Artificial stars (6 - 10 stars per image) were randomly embedded in the images, except for the edge and near the primary star. We then analyzed the data following the procedure described in \ref{analysis1} and \ref{analysis2} and repeated this process until the number of embedded stars exceeded 100 (ten or more iterations). We then averaged the self-subtractions over all data. In the area closest to the primary star ($>1\farcs5$) the amount of self-subtraction was always below 10\%.
\item{{\it Calculation of the detection limit:}\label{analysis4}} \\ We determined the $5\sigma$ detection limit by embedding stars of fixed magnitude using the SExtractor tool \citep{bertin96} to extract the point sources from the noise. Ten stars were randomly embedded at the same time in the image after ADI analysis; the number of stars extracted within $5\sigma$ was then measured. This process was repeated 10 times and the overall fraction of detected point sources was then calculated. Altering the magnitude of the artificial stars in steps of 0.1~mag, this process was repeated until the detected fraction dropped below 50\% after correcting for self-subtraction. The derived detection limits were between 17.8 and 22.1~mag, with a median of 20.5~mag.
\item{{\it Investigating characteristics of detected stars:}\label{analysis5}} \\ First, the magnitudes of detected stars were calculated. Photometry was conducted as described in procedure~\ref{analysis2}. We corrected for self-subtraction according to procedure~\ref{analysis3}. We then confirmed whether detected stars were companions or unbound stars based on results from previous studies \citep{brandt13, janson13, yamamoto13}. The confirmation involved measuring the detected stars' proper motions relative to the primary star. We used both unbound stars (background and foreground stars) as well as stars whose binary status could not be confirmed.
\end{enumerate}

Finally, the mean number of unbound stars was calculated for every field using the optimized stellar distribution model. The detection limit for each field was corrected when the predicted number was calculated. Because the signal-to-noise ratios were insufficient near the primary star and the edges of images, we removed these regions (20\% of all fields). The predicted number in the SEEDS observations was the sum of the predicted number in each field. 






%




\begin{thebibliography}{}

\bibitem[Abadi et al.(2003)]{abadi03} Abadi, M. G., Navarro, J. F., Steinmetz, M., \& Eke, V. R. 2003, \apj, 597, 21
\bibitem[Bahcall \& Soneira(1980)]{bahcall80} Bahcall, J. N., \& Soneira, R. M. 1980, \apjs, 44, 73
\bibitem[Bihain et al.(2006)]{bihain06} Bihain, G., Rebolo, R., B\'ejar, V. J. S., Caballero, J. A., Bailer-Jones, C. A. L., Mundt, R., Acosta-Pulido, J. A., Manchado Torres, A. 2006, \aap, 458, 805
\bibitem[Bensby et al.(2011)]{bensby11} Bensby, T., Alves-Brito, A., Oey, M. S., Yong, D., \& Mel\'{e}ndez, J. 2011, \apjlett, 735, 46
\bibitem[Bertin \& Arnouts(1996)]{bertin96} Bertin, E., \& Arnouts, S. 1996, \aaps, 117, 393
\bibitem[Bowler et al.(2013)]{bowler13} Bowler, B. P., Liu, M. C., Shkolnik, E. L., \& Dupuy, T. J. 2013, \apj, 774, 55
\bibitem[Bovy, Rix, and Hogg(2012)]{bovy12} Bovy, J., Rix, H.-W., \& Hogg, D. W. 2012, \apj, 751, 131
\bibitem[Brandt et al.(2013)]{brandt13} Brandt, T. D., et al. 2014, \apj, 786, 1
\bibitem[Burgasser(2007)]{burgasser07} Burgasser, A. J. 2007, \apj, 659, 655
\bibitem[Caballero et al.(2008)]{caballero08} Caballero, J. A., Burgasser, A. J., \& Klement, R. 2008, \aap, 488, 181
\bibitem[Carollo et al.(2010)]{carollo10} Carollo, D., et al. 2010, \apj, 712, 692
\bibitem[Chabriere(2003)]{chabriere03} Chabriere, G. 2003, \pasp, 115, 763
\bibitem[Chang et al.(2011)]{chang11} Chang, C.-K., Ko, C.-M., \& Peng, T.-H. 2011, \apj, 740, 34
\bibitem[Chang et al.(2012)]{chang12} Chang, C.-K., Lai, S.-Y., Ko, C.-M., \& Peng, T.-H. 2012, \apj, 759, 94
\bibitem[Chauvin et al.(2014)]{chauvin14} Chauvin, G., et al. 2014, accepted to \aap \ (arXiv:1405.1560)
\bibitem[Cruz et al.(2003)]{cruz03} Cruz, K. L., Reid, I. N., Liebert, J., Kirkpatrick, J. D, \& Lowrance, P. J. 2003, \aj, 126, 2421
\bibitem[Czekaj et al.(2014)]{czekaj14} Czekaj, M. A., Robin, A. C., Figueras, F., Luri, X., \& Haywood, M. 2014, \aap, 564, 102
\bibitem[Guo et al.(2013)]{guo13} Guo, Y., et al. 2013, \apjs, 207, 24
\bibitem[Gilmore \& Reid(1983)]{gilmore83} Gilmore, G., \& Reid, I. N. 1983, \mnras, 202, 1025
\bibitem[Holwerda et al.(2014)]{holwerda14} Holwerda, B. W., et al. 2014, \apj, 788, 77
\bibitem[Hodapp et al.(2008)]{hodapp08} Hodapp, K. W., et al. 2008, \procspie, 7014, 42
\bibitem[Hayano et al.(2010)]{hayano10} Hayano, Y., et al. 2010, \procspie, 7736, 21
\bibitem[Hayden et al.(2013)]{hayden13} Hayden, M. R., et al. 2014, \aj, 147, 116
\bibitem[Ichikawa et al.(2006)]{ichikawa06} Ichikawa, T., et al. 2006, \procspie, 6269, 38
\bibitem[Janson et al.(2013)]{janson13} Janson, M., et al. 2013, \apj, 773, 73
\bibitem[Jones et al.(1981)]{jones81} Jones, T. J., Ashley, M., Hyland, A. R., \& Ruelas-Mayorga, A. 1981, \mnras, 197, 413
\bibitem[Juri\'{c} et al.(2008)]{juric08} Juri\'{c}, M., et al. 2008, \apj, 673, 864
\bibitem[Kajisawa et al.(2011)]{kajisawa11} Kajisawa, M., et al. 2011, \pasj, 63, 379
\bibitem[Kirby et al.(2008)]{kirby08} Kirby, E. N., Simon, J. D., Geha, M., Guhathakurta, P., \& Frebel, A. 2008, \apjlett, 685, 43
\bibitem[Kirkpatrick et al.(1994)]{kirkpatrick94} Kirkpatrick, J. D., McGraw, J. T., Hess, T. R., Liebert, J., \& McCarthy, D. W. Jr. 1994, \apjs, 94, 749
\bibitem[Kirkpatrick et al.(2011)]{kirkpatrick11} Kirkpatrick, J. D., et al. 2011, \apjs, 197, 19
\bibitem[Kirkpatrick et al.(2012)]{kirkpatrick12} Kirkpatrick, J. D., et al. 2012, \apj, 753, 156
\bibitem[Konishi et al.(2013)]{konishi13} Konishi, M., Shibai, H., Matsuo, T., Yamamoto, K., Sumi, T., \& Fukagawa, M. 2013, Journal of the Japan Society of Infrared Science and Technology, 23-1, 108
\bibitem[Lafreni\`{e}re et al.(2007)]{lafreniere07} Lafreni\`{e}re, D., et al. 2007, \apj, 660, 770
\bibitem[Marois et al.(2006)]{marois06} Marois, C., Lafreni\`{e}re, D., Doyon, R., Macintosh, B., \& Nadeau, D. 2006, \apj, 641, 556
\bibitem[Nakajima et al.(2000)]{nakajima00} Nakajima, T., et al. 2000, \aj, 120, 2488
\bibitem[Nielsen et al.(2013)]{nielsen13} Nielsen, E. L., et al. 2013, \apj, 776, 4
\bibitem[Pecaut et al.(2012)]{pecaut12} Pecaut, M. J., Mamajek, E. E., \& Bubar, E. J. 2012, \apj, 746, 154
\bibitem[Pecaut \& Mamajek(2013)]{pecaut13} Pecaut, M. J, \& Mamajek, E. E. 2013, \apjs, 208, 9
\bibitem[Pirzkal et al.(2005)]{pirzkal05} Pirzkal, N., et al. 2005, \apj, 622, 319
\bibitem[Pirzkal et al.(2009)]{pirzkal09} Pirzkal, N., et al. 2009, \apj, 695, 1591
\bibitem[Polido et al.(2013)]{polido13} Polido, P., Jablonski, F., \& L\'{e}pine, J. R. D. 2013, \apj, 778, 32
\bibitem[Reid et al.(2004)]{reid04} Reid, I. N., et al. 2004, \aj, 128, 463
\bibitem[Reid et al.(2008)]{reid08} Reid, I. N., et al. 2008, \aj, 136, 1290
\bibitem[Reyl\'{e} et al.(2010)]{reyle10} Reyl\'{e}, C., et al. 2010, \aap, 522, 112
\bibitem[Ryan et al.(2011)]{ryan11} Ryan, R. E., Jr, et al. 2011, \apj, 739, 83
\bibitem[Robin et al.(2003)]{robin03} Robin, A. C., Reyl\'{e}, C., Derri\`{e}re, S., \& Picaud, S. 2003, \aap, 409, 523
\bibitem[Searle \& Zinn(1978)]{searle78} Searle, L., \& Zinn, R., 1978, \apj, 225, 357
\bibitem[Sumi et al.(2011)]{sumi11} Sumi, T., et al. 2011, \nat, 473, 349
\bibitem[Suzuki et al.(2008)]{suzuki08} Suzuki, R., et al. 2008, \pasj, 60, 1347
\bibitem[Suzuki et al.(2010)]{suzuki10} Suzuki, R., et al. 2010, \procspie, 7735, 101
\bibitem[Tamura(2009)]{tamura09} Tamura, M. 2009, in AIP Conf. Proc. 1158, Exoplanets and Disks: Their Formation and Diversity, ed. T. Usuda, M. Tamura, \& M. Ishii (Melville, NY: AIP), 11
\bibitem[Wainscoat et al.(1992)]{wainscoat92} Wainscoat, R. J., Cohen, M., Volk, K., Walker, H. J., \& Schwarts, D. E. 1992, \apjs, 83, 111
\bibitem[Windhorst et al.(2011)]{windhorst11} Windhorst, R. A., et al. 2011, \apjs, 193, 27
\bibitem[Wirth et al.(2004)]{wirth04} Wirth, G. D., et al. 2004, \aj, 127, 3121
\bibitem[Yamamoto et al.(2013)]{yamamoto13} Yamamoto, K., et al. 2013, \pasj, 65, 90   
\end{thebibliography}
\end{document}